\documentclass[]{spie}  
\usepackage[]{graphicx,bbold}
\usepackage[]{subeqn,url}
\title{Efficient analysis and representation of geophysical processes
using localized spherical basis functions}
\author{Frederik J.~Simons\supit{a}, 
Jessica C.~Hawthorne\supit{a} and Ciar{\'a}n D.~Beggan\supit{b} 
\skiplinehalf
\supit{a}
Department of Geosciences, Princeton University, Guyot Hall,
Princeton, NJ, USA\\
\supit{b}
School of GeoSciences, University of Edinburgh, Grant Institute, Edinburgh, UK.
}
\newcommand{\gaelmpc}{g^*_{\alpha\hsp lm''}}
\newcommand{\Tmmmp}{T_{m\hspace{0.125em}m'm''}}
\newcommand{\galmc}{g^*_{\alpha\hsp lm}}
\newcommand{\fpto}{s^{\phi\theta\omega}}

\newcommand{\pto}{(\phi,\theta,\omega)}
\newcommand{\gptoalmc}{\gptoc_{\alpha\hsp lm}}
\newcommand{\gptoalm}{\gpto_{\alpha\hsp lm}}
\newcommand{\gptoc}{g^{\phi\theta\omega\hsp *}}
\newcommand{\gpto}{g^{\phi\theta\omega}}
\newcommand{\rsb}{(r_s)}

\newcommand{\gaelmp}{g_{\alpha\hsp lm'}}
\newcommand{\gaelmpp}{g_{\alpha\hsp lm''}}
\newcommand{\sumshortL}{\suml_{lm}^{L}}
\newcommand{\flmd}{s^{}_{lm}}
\newcommand{\flm}{s_{lm}}
\newcommand{\bC}{\mbox{$\mathbf{C}$}}
\newcommand{\bS}{\mbox{$\mathbf{S}$}}

\newcommand{\nAlm}{{}_n\hspace{-0.05em}\bC_{lm}}

\newcommand{\nBlm}{{}_n\hspace{-0.05em}\bS_{lm}}

\newcommand{\ncAl}{{}_n\hspace{-0.05em}\mbox{\boldmath$\mathcal{A}$}_{\hspace{0.075em}l}}

\newcommand{\nol}{{}_n\hspace{-0.05em}\omega_l}

\newcommand{\udnPl}{{}^{}_n\hspace{-0.05em}P^{}_{l}}
\newcommand{\nolt}{{}^{}_n\hspace{-0.05em}\omega_l^{-2}}
\newcommand{\br}{\mathbf{r}}
\newcommand{\bM}{\mbox{$\mathbf{M}$}}

\newcommand{\sumshLp}{\suml_{l'=0}^{L}\suml_{m'=-l'}^{l'}}

\newcommand{\Drhrhp}{D(\rhat,\rhat')}

\newcommand{\domg}{\,d\Omega} 
 
\newcommand{\intr}{\int_R} 
\newcommand{\intbr}{\int_{\bar{R}}} 
 
\newcommand{\into}{\int_\Omega}

\newcommand{\also}{\quad\mbox{and}\quad} 
\newcommand{\with}{\quad\mbox{with}\quad}

\newcommand{\dab}{\delta_{\alpha\beta}}

\newcommand{\dllp}{\delta_{ll'}} 
\newcommand{\dmmp}{\delta_{mm'}} 
\newcommand{\Ylm}{Y_{lm}} 
\newcommand{\Ylmrh}{Y_{lm}(\rhat)}

\newcommand{\Ylmp}{Y_{l'm'}}

\newcommand{\Dlmlmp}{D_{lm,l'm'}} 
\newcommand{\bDlmlmp}{\bar{D}_{lm,l'm'}}

\newcommand{\grhp}{g(\rhat')}

\newcommand{\galm}{g_{\alpha\hsp lm}}

\newcommand{\glma}{g_{\alpha\hsp lm}}

\newcommand{\Lpot}{(L+1)^2} 
\newcommand{\sumapot}{\sum_{\alpha=1}^{\Lpot}} 
 
\newcommand{\sumakR}{\sum_{\alpha>N}^{\Lpot}} 
\newcommand{\sumaN}{\sum_{\alpha=1}^{N}}

\newcommand{\suml}{\sum\limits}

\newcommand{\sumsh}{\suml_{l=0}^{\infty}\suml_{m=-l}^{l}} 
\newcommand{\sumshL}{\suml_{l=0}^{L}\suml_{m=-l}^{l}}
\newcommand{\tlofp}{\left(\frac{2l+1}{4\pi}\right)} 
\newcommand{\Plm}{P_{lm}}

\newcommand{\rhat}{\mbf{\hat{r}}} 
\newcommand{\shat}{\hat{s}}

\newcommand{\be}{\begin{equation}} 
\newcommand{\ee}{\end{equation}} 
\newcommand{\ber}{\begin{eqnarray}} 
\newcommand{\eer}{\end{eqnarray}} 
\newcommand{\barray}{\begin{array}} 
\newcommand{\earray}{\end{array}} 
\newcommand{\mbf}{\mathbf}

\newcommand{\nnr}{\nonumber}

\newcommand{\rar}{\rightarrow} 
 
\newcommand{\ssec}{\subsection} 
\newcommand{\sssec}{\subsubsection} 
 
\newcommand{\hsp}{\hspace*{0.1em}}

\begin{document} 
\maketitle 

\begin{abstract}
\end{abstract}

While many geological and geophysical processes such as the melting of
icecaps, the magnetic expression of bodies emplaced in the Earth's
crust, or the surface displacement remaining after large earthquakes
are spatially localized, many of these naturally admit spectral
representations, or they may need to be extracted from data collected
globally, e.g. by satellites that circumnavigate the Earth. Wavelets
are often used to study such nonstationary processes. On the sphere,
however, many of the known constructions are somewhat limited. And in
particular, the notion of `dilation' is hard to reconcile with the
concept of a geological region with fixed boundaries being responsible
for generating the signals to be analyzed. Here, we build on our
previous work on  localized spherical analysis using an approach that
is firmly rooted in spherical harmonics. We construct, by quadratic
optimization, a set of bandlimited functions that have the majority of
their energy concentrated in an arbitrary subdomain of the unit
sphere. The `spherical Slepian basis' that results provides a
convenient way for the analysis and representation of geophysical
signals, as we show by example. We highlight the connections to
sparsity by showing that many geophysical processes are sparse in the
Slepian basis. 


\section{The spherical Slepian basis}
\label{scs}

We denote the colatitude of a geographical point $\rhat$ on the
unit sphere surface $\Omega=\{\rhat: \|\rhat\|=1\}$ by
$0\le\theta\le\pi$ and the longitude by $0\le\phi< 2\pi$. We use~$R$
to denote a region of~$\Omega$, of area~$A$, within which we seek to
concentrate a bandlimited function of position~$\rhat=(\theta,\phi)$. We use
orthonormalized \textit{real} surface spherical
harmonics~\cite{Edmonds96,Dahlen+98}, thus expressing a
square-integrable real function $f(\rhat)$ on the surface of the unit
sphere as 
\be
\label{expansion}
f(\rhat)=\sumsh {f}_{lm}\Ylmrh,\qquad {f}_{lm}=\into
f\hspace*{0.1em}\Ylm\domg,\also
\into\Ylm\Ylmp\domg=\delta_{ll'}\delta_{mm'}.
\label{normalization}
\ee
The Slepian basis for the domain~$R$ is the collection of bandlimited
functions
\be
g(\rhat)=\sumshL g_{lm}\Ylmrh
\label{bandlg}
\qquad\mbox{for which}\qquad
\lambda=\left.\intr
g^2(\rhat)\domg\right/\into^{}g^2(\rhat)\domg=\mbox{maximum} 
.
\label{normratio}
\ee
Maximizing equation~(\ref{normratio}) leads to the spectral-domain
Hermitian, positive-definite eigenvalue equation
\be
\label{fulleigen1}
\suml_{l'=0}^L\suml_{m'=-l'}^{l'}\Dlmlmp
g_{l'm'}=\lambda\hsp g_{lm},\with
\Dlmlmp=\intr\Ylm\Ylmp\domg,
\label{Dlmlmpdef}
\qquad 0\le l\le L,
\ee
but we may equally well rewrite eq.~(\ref{fulleigen1}) as a
spatial-domain eigenvalue equation: 
\be\label{firsttimeint}
\label{eig3}
\intr  \Drhrhp \,\grhp\domg'=
\lambda\hsp g(\rhat),\with
\Drhrhp
=\sum_{l=0}^L\tlofp\!P_l(\rhat\cdot\rhat'),
\qquad \rhat\in\Omega,
\label{banddelta}
\ee
where $P_l$ is the Legendre function of integer degree $l$, which
arises in this setting as a consequence of the spherical harmonic addition
theorem\cite{Edmonds96,Dahlen+98,Simons+2006a}.
Eq.~(\ref{firsttimeint}) is a homogeneous Fredholm integral equation
of the second kind, with a finite-rank, symmetric, Hermitian kernel,
and the finite set of bandlimited spatial ``Slepian'' eigensolutions
$g_1(\rhat),g_2(\rhat), \ldots,g_{(L+1)^2}(\rhat)$ is orthonormal
over the whole sphere $\Omega$ and orthogonal over the region $R$:
\be
\into g_{\alpha}g_{\beta}\domg=\delta_{\alpha\beta},\also
\intr  g_{\alpha}g_{\beta}\domg=\lambda_{\alpha}\delta_{\alpha\beta}.
\label{orthog}
\ee
When the concentration region is a circularly symmetric
cap of radius $\Theta$ centered on the North Pole the
solutions to eq.~(\ref{firsttimeint}) break down by order
$m$ and are separable in~$\theta$ and~$\phi$, and the colatitudinal
parts $g(\theta)$, which depend only on~$|m|$,
are identical to those of a
Sturm-Liouville equation which can be solved in
the spectral domain by diagonalization of a simple tridiagonal matrix
with an almost linear spectrum\cite{Simons+2006a,Simons+2007}. 
We define a space-bandwidth product or
`spherical Shannon number' by the sum of the eigenvalues,
\be
N=\sum_{\alpha =1}^{\Lpot}\lambda_{\alpha}=
\sum_{l=0}^L\sum_{m=-l}^l 
D_{lm,lm}=\int_RD(\rhat,\rhat)\,d\Omega
=\Lpot\,\frac{A}{4\pi}
.
\label{tracedef}
\ee 
The complete set of bandlimited spatial Slepian eigenfunctions
$g_1,g_2, \ldots,g_{(L+1)^2}$, irrespective of the particular region
of concentration that they were designed for, are a basis for
bandlimited scalar processes anywhere on the surface of the unit
sphere\cite{Simons+2006a,Simons+2006b}. This follows directly from the
fact that the spectral localization kernel~(\ref{Dlmlmpdef}) is real,
symmetric, and positive definite: its eigenvectors $g_{1\hsp lm},
g_{2\hsp lm}, \ldots, g_{\Lpot \hsp lm}$ form an orthogonal set, thus
the Slepian basis functions $g_\alpha(\rhat)$, $\alpha=1,\ldots,\Lpot$
given by eq.~(\ref{bandlg}) simply transform the same-sized limited
set of spherical harmonics $\Ylmrh$, $0\le l\le L$, $-l\le m\le l$
that are a basis for the same space of bandlimited spherical functions
with no power above the bandwidth $L$. After sorting the eigenvalues
in decreasing order, this transformation  orders the resulting basis
set such that the energy of the first $N$ functions, $g_{1}(\rhat),\ldots,
g_{N}(\rhat)$, with eigenvalues $\lambda\approx 1$, is
concentrated in the region~$R$, whereas the remaining eigenfunctions,
$g_{N+1}(\rhat),\ldots,g_{\Lpot}(\rhat)$, are concentrated in the
complimentary region~$\bar{R}=\Omega-R$. As in the one- and
two-dimensional case\cite{Slepian+61,Landau+61,Slepian64}, therefore,
the reduced set of basis functions  
$g_1,g_2,\ldots,g_{N}$ can be regarded as a sparse, global, basis
suitable to approximate bandlimited processes that are primarily
localized to the region~$R$. The dimensionality reduction is dependent on the
fractional area of the region of interest, i.e. the full dimension of
the space $\Lpot$ can be ``sparsified'' to an effective dimension of
$N=\Lpot A/(4\pi)$ when the signal of interest lies in a
particular geographic region. 

An example of Slepian functions on a circular region on the
surface of the sphere can be found in Figure~\ref{SPIE2009_3}.





\section{Problems in geophysics (and beyond)}

With all of the foregoing established as fact and referring again to
the literature cited so far for proof and further context, we return
to considerations closer to home, namely the estimation of geophysical
signals from noisy and incomplete observations collected at or above
the surface of the spheres ``Earth'' or ``planet''. We restrict
ourselves to real-valued scalar measurements, contaminated by
uncorrelated additive noise, which we may not know but which we shall
describe by idealized models. We focus exclusively on data acquired
and solutions expressed on the \textit{unit} sphere. We have
considered generalizations to problems involving satellite data
collected at an altitude and/or potential fields
elsewhere\cite{Wieczorek+2005,Simons+2006b,Simons+2007,Dahlen+2008}. Two
different statistical problems arise in this context, namely, (i)
how to find the ``best'' estimate of the signal given the
data\cite{Simons+2006b}, and 
(ii) how to construct from the data the ``best'' estimate of the power
spectral density of the signal in question\cite{Dahlen+2008}. In this
contribution we limit ourselves to problem~(i) as it is here that the
connections to sparsity are most readily apparent. 

Thus, let there be some data distributed on the unit sphere,
consisting of `signal', $n$ and `noise', $s$, and let there be
some region of interest $R\subset\Omega$, in other words, let
\ber
d(\rhat)=\left\{\begin{array}{ll}
s(\rhat)+n(\rhat) & \mbox{if $\rhat\in R$}\\
\mbox{unknown/undesired} & \mbox{if $\rhat\in \Omega-R$}.
\end{array}\right.
\label{datdefs}
\eer
We 
assume that the signal of interest can be expressed by way of spherical harmonic
expansion as in eq.~(\ref{expansion}), and that it is, itself, a
realization of a zero-mean, Gaussian, isotropic, random process, namely
\be\label{slmdef}
s(\rhat)=\sumsh s_{lm}\Ylmrh, \qquad
s_{lm}=\into s\,\Ylm\domg,\qquad
\langle s_{lm}\rangle =0\also 
\langle s_{lm}s_{l'm'}\rangle=S_l\,\delta_{ll'}\delta_{mm'}.
\ee
For convenience we furthermore assume that the noise is a zero-mean
stochastic process with an isotropic power spectrum, i.e. $\langle
n(\rhat)\rangle=0$ and $\langle
n_{lm}n_{l'm'}\rangle=N_l\,\delta_{ll'}\delta_{mm'}$, 
and that it is
statistically uncorrelated with the signal. We refer to 
power as \textit{white} when $S_l=S$ or $N_l=N$, or, equivalently, $\langle
n(\rhat)n(\rhat')\rangle=N\delta(\rhat,\rhat')$.
Our objective is to determine the best estimate~$\hat{s}_{lm}$ of the
spherical harmonic expansion coefficients~$s_{lm}$ of the
signal. While in the real world there can be no limit 
on bandwidth, practical restrictions force any and all of our
estimates to be bandlimited to some maximum spherical harmonic degree
$L$, thus of necessity $\hat{s}_{lm}=0$ and $\hat{S}_l=0$ when $l> L$:
\be\label{wouldbe}
\hat{s}(\rhat)=\sumshL \hat{s}_{lm}\Ylmrh
.
\ee
This limitation, combined with the statements eq.~(\ref{datdefs}) on
data coverage and the region of interest, naturally puts us back  
in the realm of `spatiospectral concentration'. As we shall
see, solving the problem at hand will gain from involving 
`localized' Slepian functions rather than, or in addition to, the
`global' spherical harmonics basis.

This leaves us to clarify what we understand by ``best'' in this
context. While we adopt the traditional statistical metrics of bias,
variance, and mean squared error to appraise the quality of our
solutions\cite{Cox+74,Bendat+2000}, the resulting connections to
sparsity will be real and immediate, owing to the Slepian functions
being naturally instrumental in constructing
efficient, consistent and/or unbiased estimates
of~$\hat{s}_{lm}$ and/or~$\hat{S}_l$. Thus, we define 
\be
v=\langle\shat^2\rangle-\langle\shat\rangle^2
,\qquad 
b=\langle\shat\rangle-s,\qquad 
\epsilon=\shat-s,
\also
\langle\epsilon^2\rangle=v+b^2
,
\ee
where the lack of subscript indicates that we can
study variance, bias and mean squared error of the estimate of the
coefficients $\hat{s}_{lm}$ but also of their spatial expansion
$\shat(\rhat)$, or indeed of their power spectrum~$\hat{S}_l$.

\ssec*{Signal estimation from noisy and incomplete
  spherical data}

\sssec*{Spherical harmonic solution}

Paraphrasing results elaborated elsewhere\cite{Simons+2006b}, we 
write the bandlimited solution to the damped inverse problem
\be 
\intr(\hat{s}-d)^2\domg+\eta\intbr \hat{s}^2\domg=\mbox{minimum}
,
\label{variational}
\ee
where $\eta\ge 0$ is a damping parameter, by straightforward algebraic
manipulation, as 
\be
\hat{s}_{lm}=\sumshLp \left(\Dlmlmp+\eta\bDlmlmp\right)^{-1} 
\intr d\,\Ylmp\domg
\label{hatss}
,
\ee
where $\bDlmlmp$, the kernel that localizes to the region $\bar{R}=\Omega-R$,
compliments $\Dlmlmp$ given by eq.~(\ref{Dlmlmpdef}) which localizes
to $R$. Given the eigenvalue spectrum of the latter, its inversion is
inherently unstable, thus eq.~(\ref{variational}) is an
ill-conditioned inverse problem unless $\eta>0$, as has been well
known, e.g. in geodesy\cite{Xu92a,Sneeuw+97}. As can be
easily shown, without damping the estimate is unbiased but effectively
incomputable; the introduction of the damping term stabilizes the
solution at the cost of added bias. And of course when $R=\Omega$,
eq.~(\ref{hatss}) is simply \textit{the}  spherical harmonic transform, as
in that case, eq.~(\ref{Dlmlmpdef}) reduces to
eq.~(\ref{normalization}), in other words, then  $\Dlmlmp=\dllp\dmmp$. 


\sssec*{Slepian basis solution}

We could seek a trial solution in the Slepian basis designed for this
region of interest~$R$ by writing
\be
\hat{s}(\rhat)=\sumapot\hat{s}_\alpha g_\alpha(\rhat) 
.
\label{shatdef1}
\ee
This would be completely equivalent to the expression in
eq.~(\ref{wouldbe}) by virtue of the completeness of the Slepian basis
for bandlimited functions everywhere on the sphere and the unitarity
of the transform~(\ref{bandlg}) from the spherical-harmonic to the
Slepian basis. The solution to the undamped ($\eta=0$) 
version of eq.~(\ref{variational}) would then be 
\be
\hat{s}_\alpha=\lambda_\alpha^{-1}\intr d\hsp g_\alpha\domg
,\label{SGsolspec}
\ee
which, being completely equivalent to eq.~(\ref{hatss}) for
$\eta=0$, would be computable, and biased, only when the expansion in
eq.~(\ref{shatdef1}) were to be truncated to some finite
$J<\Lpot$ to prevent the blowup of the eigenvalues~$\lambda$. Assuming
for simplicity of the argument that $J=N$, the 
essence of the approach is now that the solution 
\be
\hat{s}(\rhat)=\sumaN\hat{s}_\alpha g_\alpha(\rhat)
\label{shatdef2}
\ee
will be sparse (in achieving a bandwidth $L$ using $N$ Slepian
instead of $\Lpot$ spherical harmonic expansion coefficients) yet good
(in approximating the signal as best as possible in the mean squared
sense within the region of interest~$R$) and of geophysical utility
(assuming we are dealing with spatially localized processes that are
to be extracted, e.g., from global satellite
measurements)\cite{Han+2008a}. In light of the
reasoning behind eqs~(\ref{variational})--(\ref{shatdef2}), 
it is worth rereading the 1967 paper by W.~M.~Kaula\cite{Kaula67a},
which, written long before the advent of Slepian functions and the
associated mathematical machinery, paved the way for many other
studies\cite{Albertella+99,Pail+2001,Simons+2006b}.    


\ssec*{Bias and variance}

In concluding this section let us illustrate another welcome
by-product of our methodology, by writing the mean squared error for
the spherical harmonic solution~(\ref{hatss}) compared to the
equivalent expression~(\ref{SGsolspec}) for the solution in the
Slepian basis. We do this as a function of the spatial 
coordinate, in the Slepian basis for both, and, for maximum clarity of
the exposition, using the contrived case when both signal and noise
should be bandlimited as well as white (both stipulations being
technically impossible to satisfy simultaneously). In the former case,
\be 
\label{msefinal}
\langle\epsilon^2(\rhat)\rangle=N\sumapot\lambda_\alpha
[\lambda_\alpha+\eta(1-\lambda_\alpha)]^{-2}
g_\alpha^2(\rhat)+\eta^2 S\sumapot
(1-\lambda_\alpha)^2[\lambda_\alpha+\eta(1-\lambda_\alpha)]^{-2}
g_\alpha^2(\rhat)\nnr 
,
\ee
while in the latter, we obtain
\be
\langle\epsilon^2(\rhat)\rangle=N\sumaN\lambda_\alpha^{-1}g_\alpha^2(\rhat)+S\sumakR
g_\alpha^2(\rhat)
.
\label{SGmsefinal}
\ee
All $\Lpot$ basis functions are required to express the mean squared
estimation error, whether in eq.~(\ref{msefinal}) or in
eq.~(\ref{SGmsefinal}). The first term in both expressions is the
variance, which depends on the measurement noise. Without damping or truncation
the variance grows without bounds. Damping and truncation alleviate
this at the expense of added bias, which depends on the
characteristics of the signal, as given by the second term. In contrast
to eq.~(\ref{msefinal}), however, the Slepian
expression~(\ref{SGmsefinal}) has disentangled the contributions due
to noise/variance and signal/bias by projecting them onto the sparse set of
well-localized and the remaining set of poorly localized Slepian
functions, respectively. The estimation variance is felt via the
basis functions $\alpha=1\rar N$ that are well concentrated inside
the measurement area, and the effect of the bias is relegated to those
$\alpha=N+1\rar\Lpot$ functions that are confined to the region of
missing data. 

When forming a solution to our problem in the Slepian basis by
truncation according to~eq.~(\ref{shatdef2}), changing the truncation
level to values lower or higher than the Shannon number $N$
amounts to navigating the trade-off space between variance, bias (or
``resolution''), and sparsity in a manner that is captured with great
clarity by eq.~(\ref{SGmsefinal}). We refer the reader
elsewhere\cite{Simons+2006b} for more details.




\section{Examples and applications}

\ssec*{Sparsity of errors in approximation}

Suppose that instead of eq.~(\ref{variational}) we were to minimize
\be
\into(\hat{s}-d)^2\domg=\mbox{minimum}
,
\label{variational2}
\ee
which would lead to our  forming the
estimate  from the partially observed data~$d$ directly as
\be
\hat{s}_{lm}=
\intr d\,\Ylm\domg
\label{hatssv}
.
\ee 
While this might seem to be the natural approach in 
the absence of information outside the region of interest~$R$, this would be
equivalent to solving eq.~(\ref{variational}) using a non-optimal damping
constraint of $\eta=1$, as we have noted
elsewhere\cite{Simons+2006b}, and eq.~(\ref{hatss}) furthermore shows this
is so since $\Dlmlmp+\bDlmlmp=\dllp\dmmp$. Nevertheless, we use this
example because it is simple and informative, and it has been studied by
others before\cite{Sneeuw+97,VanGelderen+97}. If, for simplicity we
assume that both signal and noise have a white power spectrum, the
spectral error covariance matrix is
\be\label{errco1}
\langle\epsilon_{lm}\epsilon_{l'm'}\rangle=N\Dlmlmp+S\bDlmlmp
,
\ee
as can be derived by combining eq.~(\ref{hatssv}) with
eqs~(\ref{Dlmlmpdef}) and~(\ref{datdefs})--(\ref{slmdef}). The errors
are correlated, and are due to noise  
in the region~$R$ where data exist, and contaminated by signal from the
missing region~$\bar{R}=\Omega-R$. It now makes immediate intuitive sense to
desire a parameterization whose estimated coefficients have a
diagonal covariance matrix. Let this parameterization be in terms of
Slepian functions and thus this estimator 
\be
\hat{s}_\alpha=\intr d\hsp g_\alpha\domg
,\label{SGsolspecv}
\ee
which is to be contrasted with eq.~(\ref{hatssv}), and the
associated error covariance will be diagonal and given by
\be
\langle\label{errco2}
\epsilon_\alpha\epsilon_\beta\rangle=[N\lambda_\alpha+S(1-\lambda_\alpha)]
\,\dab
.
\ee
Eq.~(\ref{errco2}) follows from eq.~(\ref{errco1}) by the unitarity of the
transform~(\ref{bandlg}) and the properties~(\ref{normalization})
and~(\ref{fulleigen1}). 

\begin{figure}\center
\includegraphics[width=0.75\columnwidth]{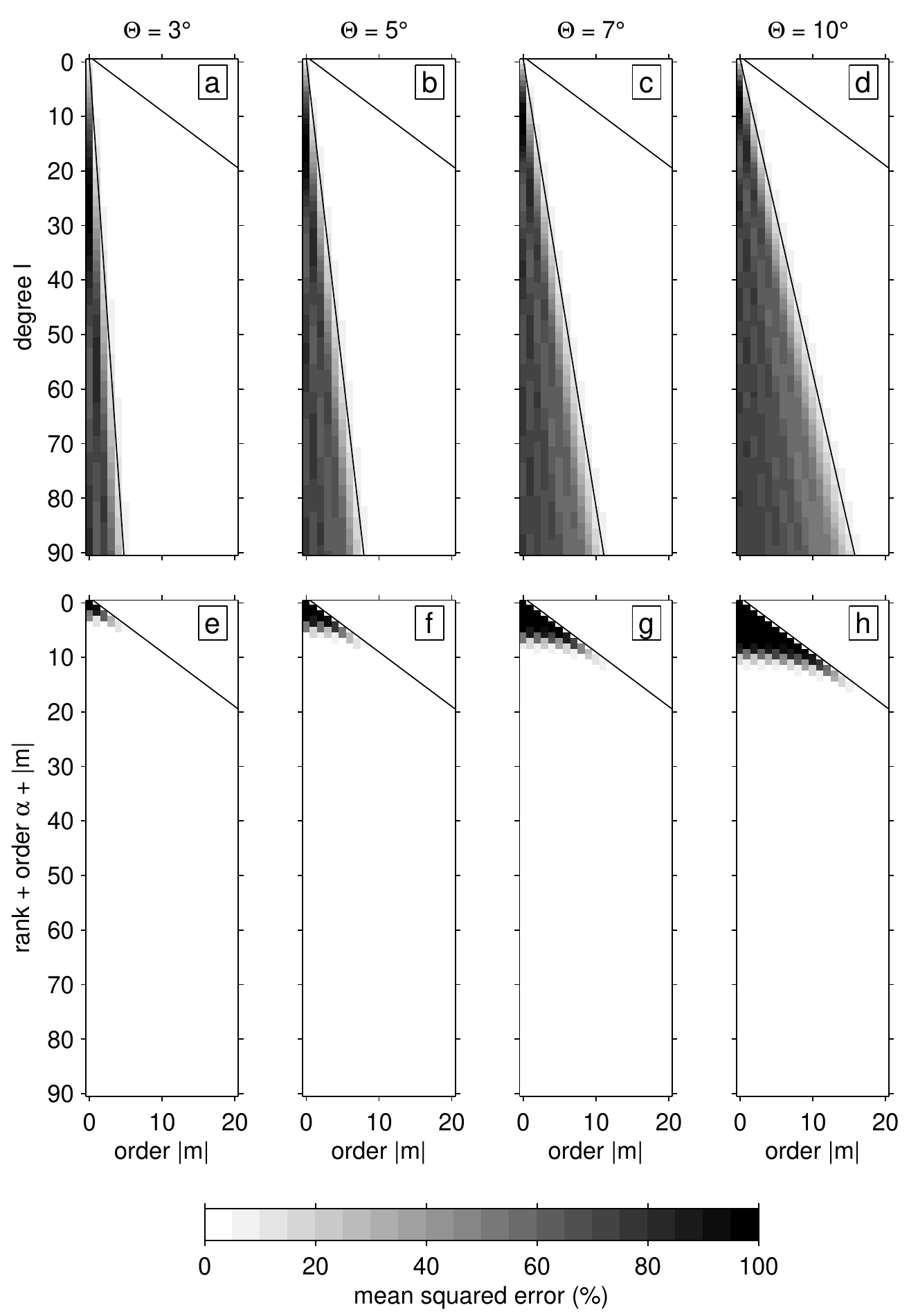}
\caption{\label{SPIE2009_2}Error covariance of the solutions to the
geodetic estimation problem for scalar data observed in a region~$R$
in the presence of a `polar gap', an axisymmetric pair of antipodal
polar caps $\bar{R}=\Omega-R$ of radii
$\Theta=0^\circ,3^\circ,5^\circ,7^\circ$, and~$10^\circ$, as shown.
(\textit{a}--\textit{d}) Diagonal elements of the error covariance
matrix, eq.~(\ref{errco1}), of the spherical harmonic
solution~(\ref{hatssv}).  (\textit{e}--\textit{h}) Diagonal elements
of the error covariance matrix, eq.~(\ref{errco2}), of the $L=90$
Slepian basis solution~(\ref{SGsolspecv}). The ordinate is the sum of
the rank~$\alpha$ of the Slepian function within a sequence of single
absolute order and this order,~$|m|$. The lines at $l=|m|$ in panels
\textit{a}--\textit{h} mark the area of the space of spherical
harmonics at degrees $l=0,\ldots, L$, and orders $|m|=0,\ldots, l$,
while the lines at $|m|=(l+1/2)\sin\Theta_0$ in panels
\textit{a}--\textit{d} delineate the approximate influence zone of the
errors.}
\end{figure}

An example of this behavior can be found in Figure~\ref{SPIE2009_2},
where the mean squared errors $\langle\epsilon_{lm}^2\rangle$ and $\langle
\epsilon_\alpha^2\rangle$ are plotted for the noiseless case,
$N=0$, as a percentage of the signal strength, $S$. The geometry is that of a
typical satellite survey characterized by a `polar gap'
$\bar{R}=\Omega-R$ consisting of a pair of
axisymmetric caps at the North and South Pole, respectively, each one
of various colatitudinal radii
$\Theta=0^\circ,3^\circ,5^\circ,7^\circ$, and $10^\circ$. The
 wedge shape of the errors in
Figure~\ref{SPIE2009_2}\textit{a}--\textit{d} is bounded by the line
$|m|=(l+1/2)\sin\Theta_0$. This relation substantiates earlier, 
heuristic, findings\cite{VanGelderen+97,Sneeuw+97}: it identifies
$\Theta_0$ as the `turning colatitude'\cite{Dahlen+98} separating the
oscillatory from the evanescent parts of an asymptotic approximation
of the Legendre functions at a given degree~$l$ and
order~$m$. Compared to the error structure of the spherical harmonic
solution~(\ref{hatssv}) shown in
Figure~\ref{SPIE2009_2}\textit{a}--\textit{d}, the error structure
of the Slepian basis solution~(\ref{SGsolspecv}) shown in
Figure~\ref{SPIE2009_2}\textit{e}--\textit{h} is decidedly more sparse. 


\ssec*{Sparsity from geometry}

Geophysical signals that are regional in nature are sparse in the
Slepian domain --- provided the Slepian basis constructed is
commensurate with the localization of the signal itself. To illustrate
this we focus no longer on estimating the field but rather on the
representation and approximation of the solution once 
it has been obtained. To this end we will drop the hats, omit the
angular brackets on all symbols, and ignore observational
noise.

A bandlimited signal can be represented equally well in the
spherical-harmonic as in any kind of Slepian basis of the same
bandlimitation. When the signal is local and the chosen Slepian basis
is similarly localized, approximating the function by the expansion
truncated at the Shannon number will be advantageous
\be
s(\rhat)=
\sumshL s_{lm}\Ylmrh=
\sumapot s_\alpha g_\alpha(\rhat) 
\approx\sumaN s_\alpha g_\alpha(\rhat)
,
\label{shatdef3}
\ee
both because the quality of the approximation will be high in the
region of interest and because the Shannon number of  terms
contributing to the reconstruction will be a
significant savings over the full number of expansion
coefficients. As per eq.~(\ref{tracedef}) this sparsity is thus mostly
``geometric'' in origin and the efficiency gains are dependent on the
area of the region of interest expressed as a fraction of the area of
the entire unit sphere.

Using the double orthogonality definitions~(\ref{orthog}) it is easy
to derive from eq.~(\ref{shatdef3}) that the mean squared error (mse) over
the region of interest of such an approximation will depend on the signal 
terms neglected in the expansion, and that the ``$R$-average mse'' as
a fraction of the $R$-average mean signal strength should be given by
\be
\left.\intr \epsilon^2(\rhat)\domg\right/
\intr s^2(\rhat)\domg=
\left.\sumakR s_\alpha^2\lambda_\alpha\right/
\sumapot s_\alpha^2\lambda_\alpha
.
\ee
The quality of the regional approximation is high, because
$\lambda_\alpha\approx 1$ when $\alpha\le N$ and
$\lambda_\alpha\approx 0$ when $\alpha> N$. 

\begin{figure}\center
\includegraphics[width=0.9\columnwidth]{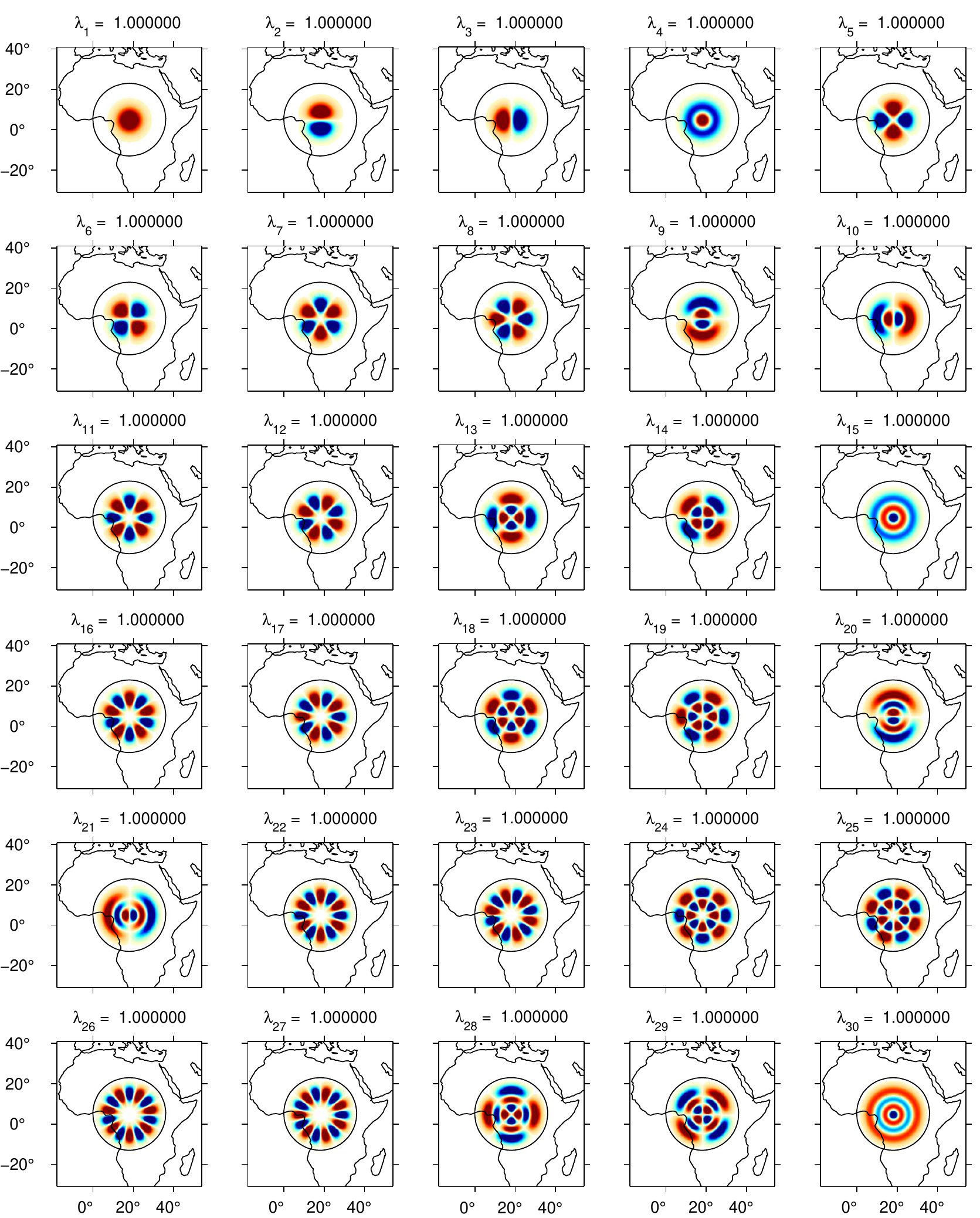}
\caption{\label{SPIE2009_3}Bandlimited eigenfunctions $g(\theta,\phi)$
that are optimally concentrated within a circularly symmetric domain
of colatitudinal radius $\Theta=18^\circ$ centered on
$\theta_0=85^\circ$ and $\phi_0=18^\circ$. The bandwidth is $L=72$ and
the rounded Shannon number $N=130$. The circle denotes the cap
boundary. Blue is positive and red is negative and the color axis is
symmetric, but the sign is arbitrary; regions in which the absolute
value is less than one hundredth of the maximum value on the sphere
are left white.}
\end{figure}

\begin{figure}\center
\includegraphics[width=0.9\columnwidth]{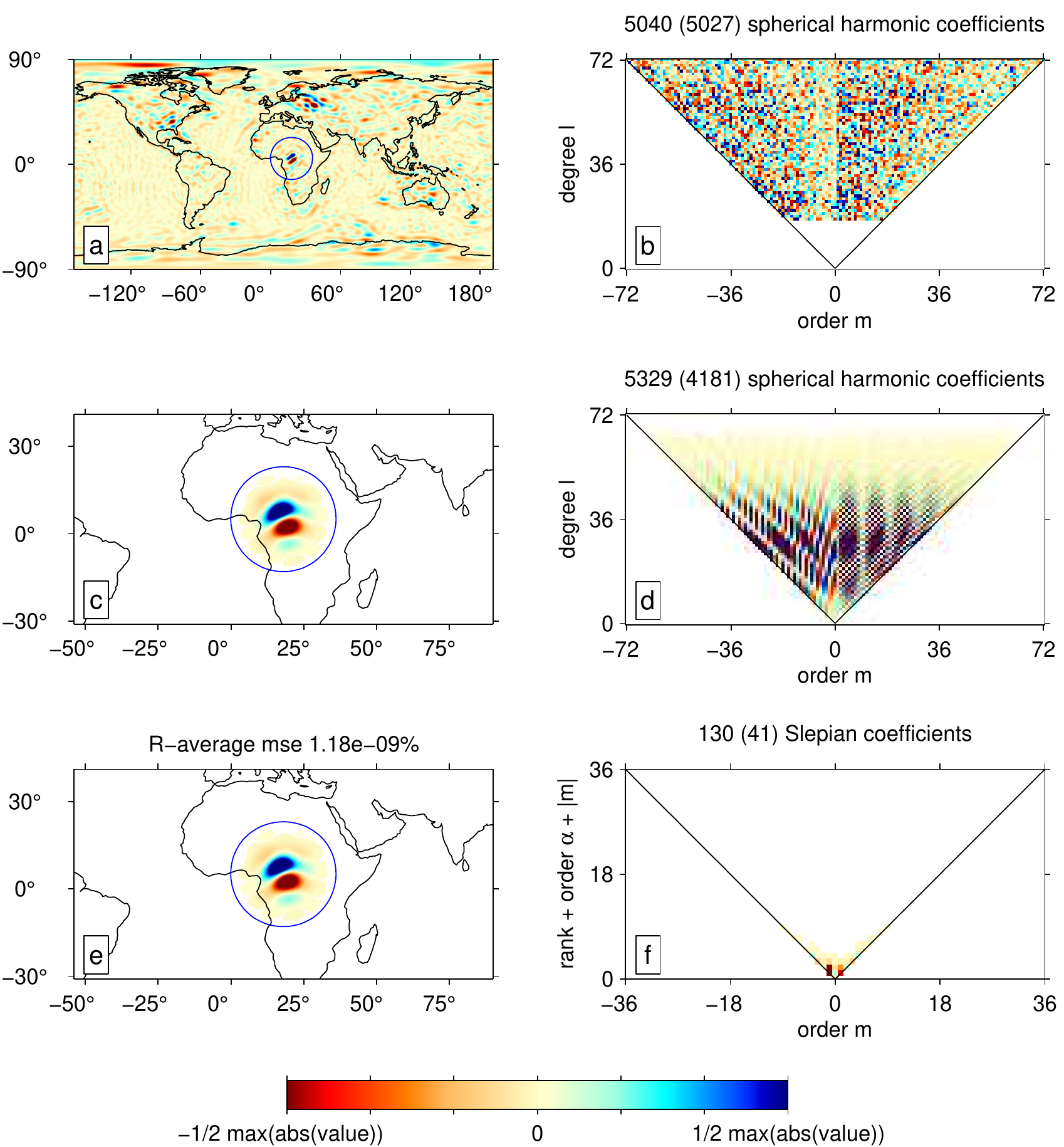}
\caption{\label{sparsity}
Global and local representations of the lithospheric magnetic
field in the spherical-harmonic and Slepian bases.
(\textit{a})
Map of the radial component of the internal magnetic field at the
Earth's surface according to version 4.2s of the POMME
model\cite{Maus+2006}, bandpassed between spherical harmonic degrees
17 and 72, and 
(\textit{b}) the spherical harmonic coefficients themselves. In total
5040 coefficients are needed to represent the global field, of which
5027 exceed a ``1/1000'' significance threshold of one thousandth of
the maximum absolute value of all coefficients. Values below this
relative threshold are left white in this and all other panels.
(\textit{c}) Map of what is known as the ``Bangui anomaly'', a highly
localized feature in central Africa. The anomaly was obtained by
multiplying the global field, low-passed to degree 36, by the Slepian
function of bandwidth 36 that is best concentrated to the circular
area of radius $\Theta=18^{\circ}$, shown in blue. The resulting
localized field is now 
bandlimited to degree 72, as can be seen in (\textit{d}). Of the 5329
coefficients necessary to represent this anomaly, 4181 exceed the
``1/1000'' threshold. 
(\textit{e}) An approximation of the same anomaly using the
$N=130$ best-localized of the 5329 Slepian functions concentrated
to the region and bandlimited to degree 72. Of the 130  
coefficients only 41 exceed the ``1/1000'' threshold,  as shown in
(\textit{f}), which has a truncated ordinate. The approximation in the
region  of interest is beyond reproach and the representation by the Slepian
expansion, compared to the spherical harmonics, is truly sparse.
} 
\end{figure}

Figure~\ref{sparsity} illustrates this by expanding the regional
``Bangui anomaly'' in the lithospheric magnetic field in both the
spherical-harmonic and Slepian bases. As can be seen in
Figure~\ref{sparsity}\textit{a}--\textit{b}, the radial component of
the Earth's main field, shown according to the POMME
model\cite{Maus+2006} bandpass filtered to between degrees 17 and 72,
contains some very prominent energy near Bangui, the capital of the
Central African Republic. The origin of this anomaly remains
debated\cite{Gubbins+2007}. Windowing a spatial expansion of the field
between degrees 17 and 36 with a Slepian window of~$L=36$ concentrated
to a circular patch of radius~$\Theta=18^{\circ}$ results in the
rendition shown in Figure~\ref{sparsity}\textit{c}--\textit{d}. In the
spherical harmonic domain, the windowed anomaly now contains energy
between degrees~$l=0-72$, as can be easily
derived\cite{Wieczorek+2005}. Almost 80\% of the spherical harmonic
expansion coefficients are significant in that their absolute value
rises above a threshold of one thousandth of their maximum absolute
value. Switching bases and expanding the signal in the $L=72$ Slepian
basis localized to the same $\Theta=18^{\circ}$ region (i.e. those
portrayed in Figure~\ref{SPIE2009_3}) results in a very small number
of significant expansion coefficients. Indeed, the partial
reconstruction using the first $N=130$ basis functions reveals
that the anomaly is extremely well captured inside of the region of
interest while the contribution to the signal is comprised of the
energy of a mere 41 coefficients that are significant according to the
same criterion, as shown in
Figure~\ref{sparsity}\textit{e}--\textit{f}.

\ssec*{Sparsity from (geo)physics}

It has long been known that earthquakes perturb the terrestrial
gravity field\cite{Chao+87}. Recently, thanks to the time-variable
gravity measurements of the GRACE satellite pair\cite{Tapley+2004a},
the study of such changes has received renewed 
attention\cite{Han+2006b,Han+2008a,Han+2008c,Linage+2009}. Following
seismological convention\cite{Dahlen+98}, let us denote the
\textit{hypo}center of an earthquake, which is 
to be considered as a point source, as~$\br_s=(r_s,\theta_s,\phi_s)$, and let
us vectorize the symmetric `moment tensor' as
\be
\label{momtens}
\bM=\left[\!\!
\barray{cccccc}
M_{rr} & M_{\theta\theta}& M_{\phi\phi}&
M_{r\theta}& M_{r\phi}& M_{\theta\phi}
\earray
\!\!\right]
.
\ee
In a coordinate system~$\br'=(r,\theta',\phi')$ that is centered on
the \textit{epi}center of the earthquake, the first-order Eulerian
gravitational potential perturbation in a spherically-symmetric
non-rotating Earth is given by a sum over `spheroidal normal-mode'
eigenfunctions~$\udnPl(r)$ (which depend on the 
Earth model, in our case \texttt{prem}\cite{Dziewonski+81a}),
\be
\label{phiE1epi}
\phi^{\mathrm{E}1}(\br')=
\bM\cdot\sum_{n=0}^{\infty}\sum_{l=0}^{\infty}
\nolt \udnPl(r)
\left(\frac{2l+1}{4\pi}\right)
\ncAl(r_s,\theta',\phi')
,
\ee
where~$l$ is the usual spherical harmonic degree, $\nol$ the temporal frequency,
and the vector excitation amplitude 
\be\label{vecA}
\ncAl(r_s,\theta',\phi')=
\sum_{m=0}^{\mathrm{min}(2,l)}
\left[
\nAlm\rsb\cos m\phi'+
\nBlm\rsb\sin m\phi'
\right] \Plm(\cos\theta')
,
\ee
with~$m$ the usual spherical harmonic order, and with the associated
Legendre function, $\Plm$, evaluated at the epicentral angular
distance~$\theta'$.  The vector functions $\nAlm$ and $\nBlm$ are
combinations of normal-mode displacement eigenfunctions and their
radial derivatives\cite{Dahlen+98,Nissen-Meyer+2007a}, and need to be
precomputed in the Earth model of choice. 

\begin{figure}\center
\includegraphics[width=0.7\columnwidth]{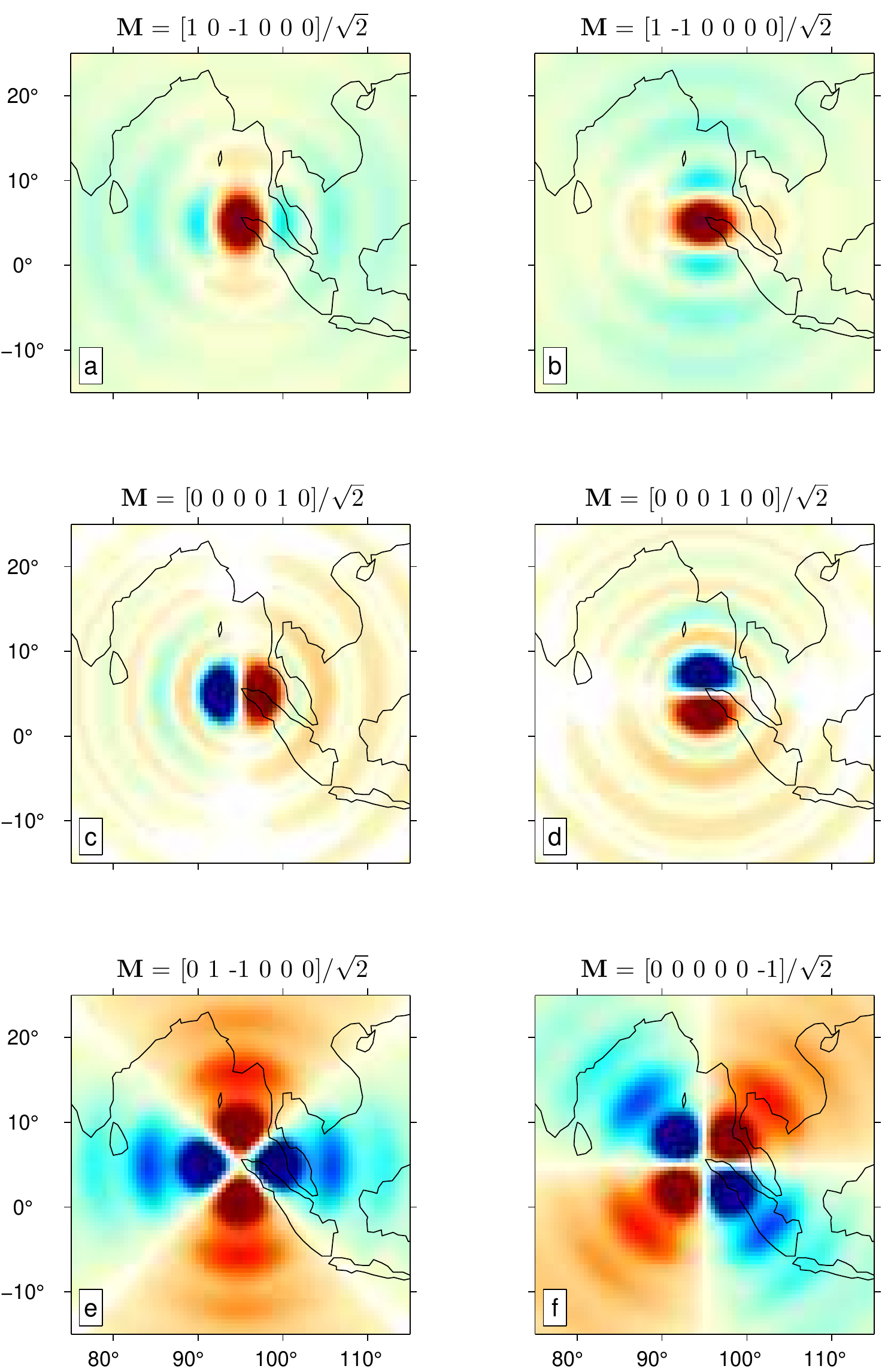}
\caption{\label{premquakes}The patterns of gravitational potential
perturbation owing to fictitious and idealized deviatoric
double-couple point-source earthquakes occurring at 30~km depth
underneath the island of Sumatra in the spherically symmetric Earth
model \texttt{prem}\cite{Dziewonski+81a}, calculated from normal-mode
theory\cite{Dahlen+98} complete to degree~$L=60$ . Blue is positive
and red is negative; the color axes are symmetric. The components of
the moment tensor, eq.~(\ref{momtens}), of these end-member cases are
indicated by the title.  (\textit{a})--(\textit{b}) 45$^\circ$-dip
thrust faults.  (\textit{c})--(\textit{d}) Vertical dip-slip faults.
(\textit{e})--(\textit{f}) Vertical strike-slip faults. All focal
mechanisms were normalized to unit scalar moment magnitude. The
relative proportions of the magnitudes of the response shown in each
row are 100\%, 54\%, and 34\%, respectively.
}
\end{figure}

Solutions~(\ref{phiE1epi}) for a variety of
end-member earthquake sources are shown in Figure~\ref{premquakes}.
As suggested by the summation limits of
eq.~(\ref{vecA}), the patterns with which earthquakes perturb the
Earth's gravity field have the symmetries of monopoles, dipoles, and
quadrupoles\cite{Dahlen+98,Nissen-Meyer+2007a}. They should thus be
eminently suitable to representation and amenable to analysis in the
Slepian basis, in which, in other words, this type of geophysical
signal is sparse. 

To the standard treatment using spherical Slepian functions we add one
sophistication, namely rotation about their center of
figure. As we have noted the Slepian basis set on
circularly symmetric domains is degenerate and separable into
`colatitudinal' eigenfunctions which are solutions to a
Sturm-Liouville equation that can be solved spectrally with great
ease\cite{Simons+2006a,Simons+2006b,Simons+2007}, and 
order-dependent `longitudinal' functions that control the axial
symmetry. To compute the functions shown in Figure~\ref{SPIE2009_3} we
first determined the spherical harmonic expansion coefficients of the
solutions to eq.~(\ref{fulleigen1}) centered on the North Pole, and
subsequently rotated those to the colatitude and longitude of the
desired cap center. Now we shall rotate the resulting functions over a
third Euler angle about their center. 

We begin by a bait-and-switch for convenience --- compared to
Section~\ref{scs} we will now work in the basis of \textit{complex}
surface spherical harmonics\cite{Edmonds96,Dahlen+98}, according to which the real 
Slepian and complex spherical harmonic expansion
coefficients of the real signal~$s(\rhat)$, $s_\alpha$ and $\flm$,
respectively,  are related via the unitary transform 
\be
s_\alpha=\sumshortL\galmc \flm^{}
\also
\flm=\sumapot\glma s_{\alpha}
\label{rota}
,
\ee
where the $\galm$ form an $\Lpot\times\Lpot$ \textit{complex}
matrix. We shall here be concerned with
obtaining the `localized' coefficients~$s_\alpha$ only when the
spherical harmonic expansion coefficients $\flm$ are ``known'',
e.g. by being given in the form of so-called `Level-2' GRACE data
products\cite{Han+2007,Han+2008a}. Alternatives to this approach that
can work directly from the raw data collected are discussed
elsewhere\cite{Simons+2006b,Han2008,Han+2008c}. 


Next, we adorn the Slepian basis functions, $g_\alpha$, their
spherical harmonic expansion coefficients, $\galm$, and the
Slepian expansion coefficients of the signal, $s_\alpha$, by three
upper indices, $\omega$, $\theta$ and $\phi$, describing,
respectively, the rotation about their center of symmetry,
$0\le\omega<2\pi$, and its colatitude, $0\le\theta\le\pi$, and
longitude, $0\le\phi<2\pi$. Thus, the ``mother'' Slepian
basis set, concentrated over the circular region~$R$ centered on the
North Pole, bandlimited to~$L$, and of Shannon number~$N$, consists of
the functions~$g^{000}_\alpha(\rhat)$; they spawn a family of functions
$\gpto_{\alpha}(\rhat)$, $\alpha=1,\dots,\Lpot$, centered at
the geographical locations~$(\theta,\phi)$ on the unit sphere and
rotated by~$\omega$. For convenience we write no superscripts when 
$\pto=(0,0,0)$. The triad $\pto$ contains the three Euler
angles that generate a Slepian basis of arbitrary orientation
anywhere on the sphere: first, by rotation over~$\omega$ around the
$z$ axis, then by~$\theta$ about the original~$y$, and finally
by~$\phi$ around the original~$z$ axis again. The spherical harmonic
coefficients of  the rotated~$\gpto_{\alpha}$ and the mother
set~$g_{\alpha}$ are then related by\cite{Edmonds96,Dahlen+98} 
\be
\gptoalm=
\sum_{m'=-l}^l\mathcal{D}^{(l)}_{mm'}\pto\,\gaelmp
,\label{grot}
\ee
where $\mathcal{D}^{(l)}_{mm'}\pto$ is a $(2l+1)\times(2l+1)$ unitary
transformation matrix of `generalized' Legendre
functions,\cite{Risbo96,Choi+99} 
\be
\mathcal{D}^{(l)}_{mm'}\pto=
e^{im\omega}d^{(l)}_{mm'}(\theta)\,e^{im'\phi}
,\label{dlmb}
\ee
duly noting that $d^{(l)}_{mm'}(\theta)=d^{(l)}_{m'm}(-\theta)$. It is
now convenient\cite{Risbo96,Wandelt+2001b} to decompose a  
general rotation to $\pto$ into one over $(\phi-\pi/2,-\pi/2,\theta)$  
followed by another over $(0,\pi/2,\omega+\pi/2)$. In that case eq.~(\ref{grot}) can
be rewritten as 
\be\label{gptoalm}
\gptoalm=
\sum_{m'=-l}^l
\sum_{m''=-l}^l
d^{(l)}_{mm'}\!\left(\frac{\pi}{2}\right)
d^{(l)}_{m''m'}\!\left(\frac{\pi}{2}\right)
e^{im(\omega+\pi/2)}
e^{im'\theta}
e^{im''(\phi-\pi/2)}
\,\gaelmpp
.
\ee
The expansion of the signal~$s(\rhat)$ into the rotated
Slepian basis $\gpto_\alpha(\rhat)$ is given, as in
eqs~(\ref{shatdef1}) and~(\ref{rota}), by 
\be\label{fexpand2}
s(\rhat)=\sumapot \fpto_\alpha \,\gpto_\alpha(\rhat)\also
\fpto_\alpha=\sumshortL\gptoalmc \flmd
.
\ee
Combining eqs~(\ref{gptoalm})--(\ref{fexpand2}) and rearranging the
summations we can write the `fast Slepian expansion' as a discrete
Fourier transform\cite{Wandelt+2001b,McEwen+2007}, to be calculated on
a grid of frequencies by FFT of the operator~$\Tmmmp^\alpha$,
\begin{subequations}
\be
\fpto_\alpha=
\sum_{m=-L}^L
\sum_{m'=-L}^L
\sum_{m''=-L}^L
\Tmmmp^\alpha\,
e^{-im\omega}
e^{-im'\theta}
e^{-im''\phi}
,
\ee
\be
\Tmmmp^\alpha=
\sum_{l=\max(|m|,|m'|,|m''|)}^{L}
d^{(l)}_{mm'}\!\left(\frac{\pi}{2}\right)
d^{(l)}_{m''m'}\!\left(\frac{\pi}{2}\right)
e^{i(m''-m)\pi/2}
\,\gaelmpc
\flm.
\ee
\end{subequations}

\begin{figure}\center
\rotatebox{-90}{\includegraphics[height=1\columnwidth]{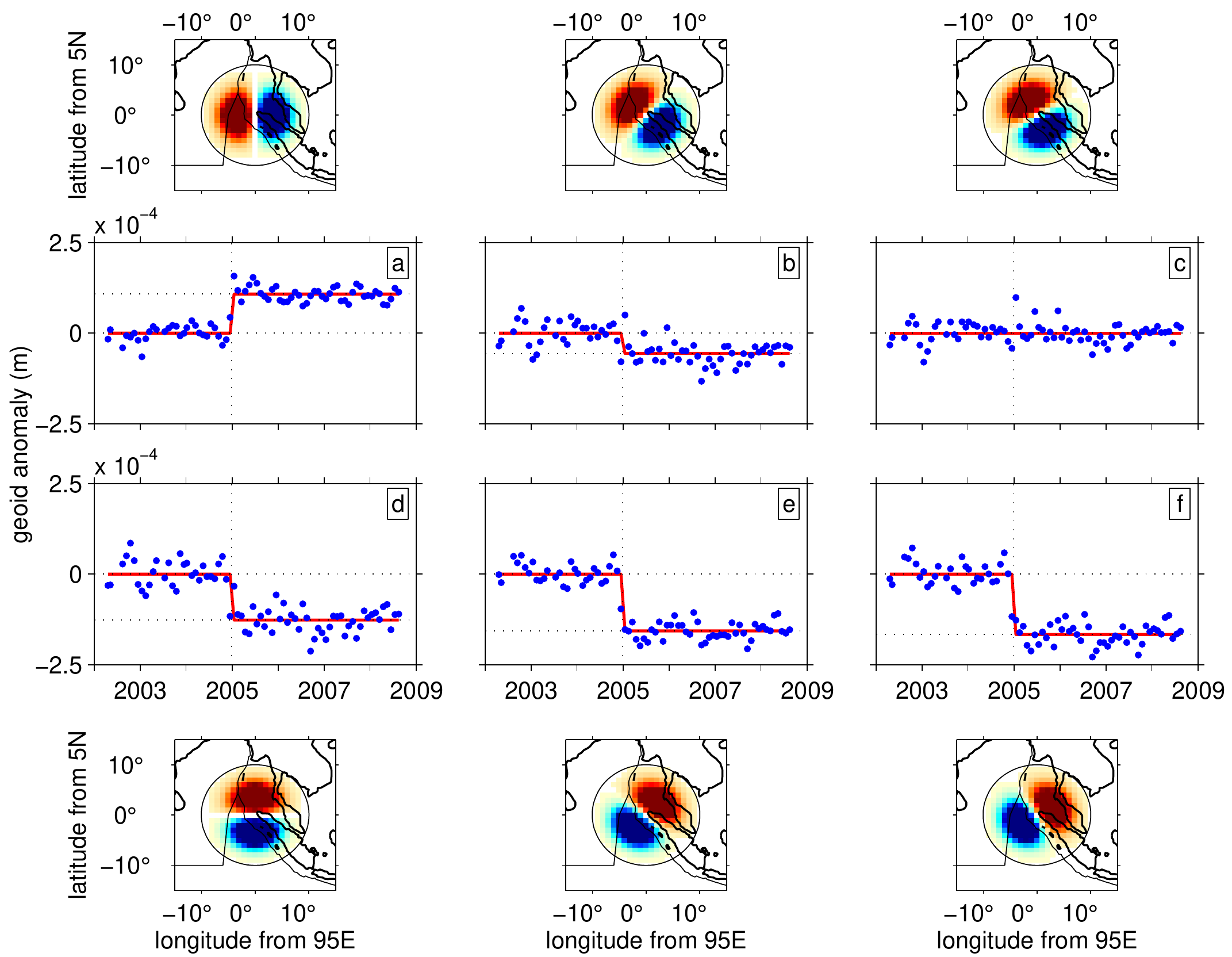}}
\caption{\label{jessica1}Orientation selectivity of the Slepian basis
functions and the signal from the 12/26/2004 Sumatra-Andaman
earthquake. The `Level-2' time-variable gravity spherical harmonic
coefficients from the Gravity Recovery And Climate
Experiment\cite{Tapley+2004a} (GRACE) were transformed via
eq.~(\ref{fexpand2}) to the expansion coefficients in a circularly
symmetric Slepian basis of colatitudinal radius~$\Theta=10^\circ$ and
of bandwidth~$L=60$, centered on the northwestern tip of the island of
Sumatra, $\theta_0=85^\circ$ and $\phi_0=95^\circ$. A linear trend,
annual and semiannual variations were removed prior to display. The
monthly varying contributions from the $m=\pm 1$  best-concentrated
($\alpha=1$) basis functions are shown after rotation of the basis
over the angles~$\omega=0^\circ, -30^\circ$ and~$-50^\circ$,
respectively. The final rotation projects almost all of the energy of
the signal onto a single component. By this measure, the best-fitting
estimate of the overall strike of the earthquake is N$40^\circ$W,
which is in good agreement with independent seismological observations
of this complex rupture\cite{Lay+2005}.
}
\end{figure}

Figure~\ref{jessica1} applies this procedure to the time-variable
gravity field as seen by GRACE for the Slepian basis of colatitudinal
radius~$\Theta=10^\circ$ and bandwidth~$L=60$ centered at
$\theta_0=85^\circ$ and $\phi_0=95^\circ$ and for varying
rotations~$\omega$. The gravitational-potential perturbation due to
the 2004 Sumatra-Andaman earthquake is a clearly visible step in the
time series of the expansion coefficients belonging to the $\alpha=1$
best concentrated $m=\pm 1$ Slepian functions, expressed as
height anomalies in~m above the Earth's reference geoid. The
orientation of the best-fitting fault plane can be derived from their
relative contributions. 

\begin{figure}\center
\includegraphics[width=1\columnwidth]{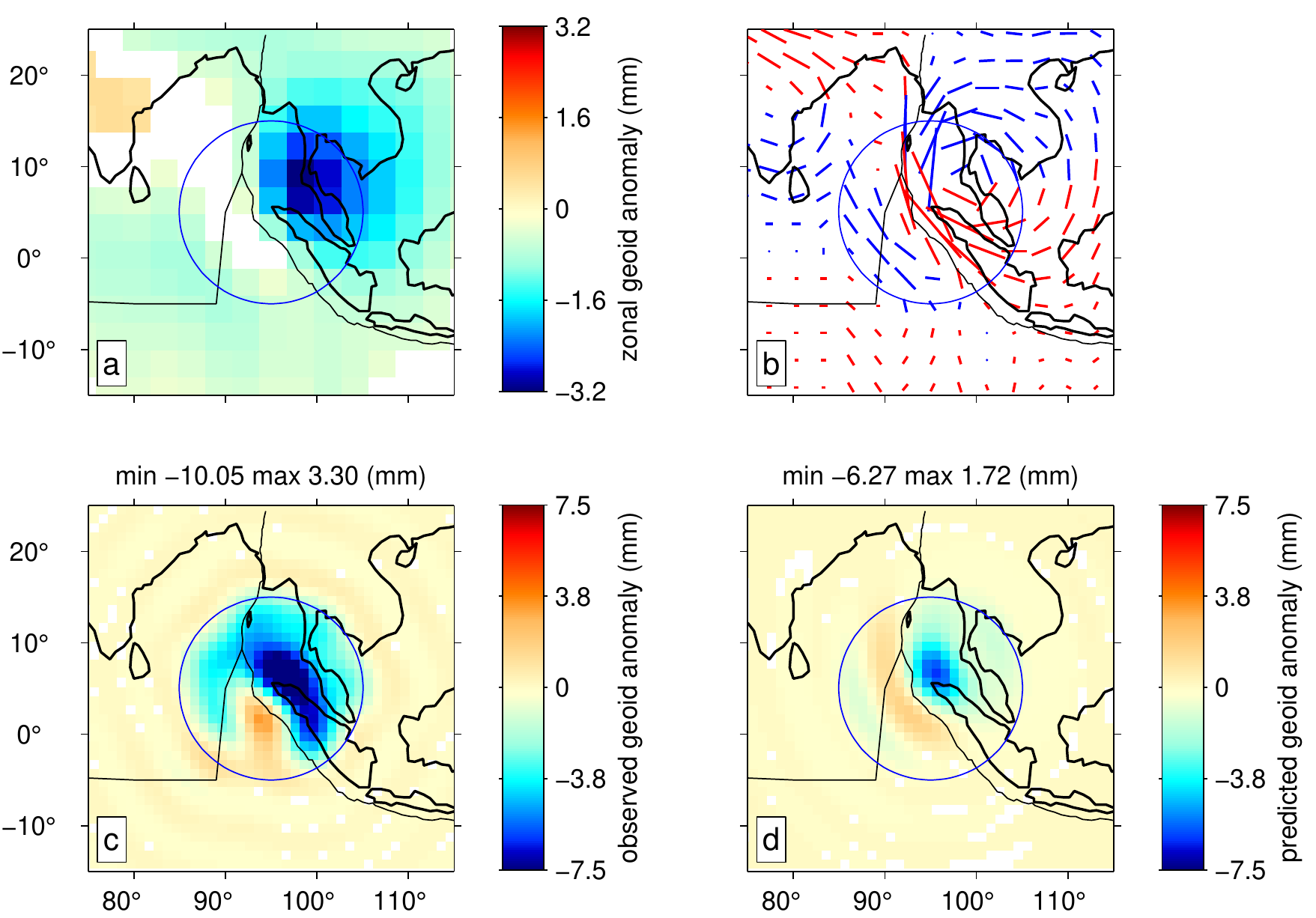}
\caption{\label{jessica2}The signal from the 12/26/2004
Sumatra-Andaman earthquake. The analysis was carried out using
identical monthly solutions from GRACE as in Figure~\ref{jessica1} and with
the same Slepian basis functions, for which $\Theta=10^\circ$ and
$L=60$, but with their centers shifted as discussed below. The
processing was identical in that a linear trend and (semi)annual
variations were fitted and removed before inverting the resulting time
series for the step-change due to the earthquake (the ``coseismic
signal''). (\textit{a}) Magnitude of the zonal signal, i.e. the
pointwise spatial expansion of the~$m=0$ Slepian
coefficients of the geoidal perturbation obtained after inversion of
the time series for a step increase or decrease, using different
Slepian basis sets whose 
centers $(\theta,\phi)$ coincide with the pixels being shown.
(\textit{b}) Directionality of the earthquake signal as measured by
the pointwise expansion of the vectorial magnitude and direction of
the $m=\pm 1$, $\alpha=1$ Slepian coefficients of the geoid change obtained by
inversion of the GRACE time series. Every arrow drawn corresponds to
the solution in a different Slepian basis centered at
the current location, as is the case for each pixel in
Figure~\ref{jessica2}\textit{a}. Blue and red indicate that the
positive and negative parts, respectively, of the best-fitting rotated
Slepian function are to be found to the north of the arrow.
(\textit{c}) The best estimate of the coseismic geoid change due to
the Sumatra-Andaman earthquake obtained from inversion of the Slepian
expansion of the GRACE time series for a step change at the known
earthquake time, also allowing for an exponential relaxation
of the perturbation (the ``postseismic
signal''\cite{Han+2008c,Linage+2009}). The spatial 
expansion is complete: the single Slepian basis used was centered on
$\theta_0=85^\circ$ and $\phi_0=95^\circ$. (\textit{d}) Spatial
rendition of the geoid change predicted independently, from seismic and
geodynamical modeling of the
earthquake\cite{Han+2006b,Han+2008a}. This forward model is based on
several simplifying assumptions\cite{Han+2006b}, and more research is
needed in order to compare
Figures~\ref{jessica2}\textit{c}--\textit{d} geophysically.
}
\end{figure}

Figure~\ref{jessica2}, finally, shows our best overall estimate of the
geoid perturbation due to the Sumatra-Andaman
earthquake. Figure~\ref{jessica2}\textit{a}--\textit{b} show
renderings of the signal projected onto the best-concentrated $m=0$
and $m=\pm 1$ components, respectively, of a Slepian basis
with~$\Theta=10^\circ$ and~$L=60$ and varying center locations. Each
pixel in Figure~\ref{jessica2}\textit{a} and each 
arrow in Figure~\ref{jessica2}\textit{b} plots the results
from a different Slepian basis centered at the applicable
locations~$(\theta,\phi)$ shown. Figure~\ref{jessica2}\textit{c} shows
the results from an inversion for the statistically significant
``coseismic'' step-changes of the geoid due to the earthquake in each
of the well-concentrated basis functions of the single Slepian basis
with~$\Theta=10^\circ$ and~$L=60$ centered at $\theta_0=85^\circ$ and
$\phi_0=95^\circ$, allowing also for exponential ``postseismic''
relaxation\cite{Han+2008c,Linage+2009}.
Figure~\ref{jessica2}\textit{d} shows the signal predicted
independently by expanding the geoid change obtained from a combined  
seismological-geodynamical forward model of the earthquake
rupture and the associated coseismic perturbations in a simplified
Cartesian Earth model\cite{Han+2006b,Han+2008a}. More ample discussion
of these models and their differences will take place in the
geophysical literature. 

\acknowledgments

Financial support for this work has been provided by the
U.~S.~National Science Foundation under Grants EAR-0105387 and
EAR-0710860. We thank Shin-Chan Han, Mark Panning and Mark Wieczorek
for discussions, and Yann Capdeville for making his normal-mode
eigenfunction code \texttt{minosy} publicly available. Computer
algorithms are made available on \url{www.frederik.net}.


\newpage

\newpage

\newpage

\newpage

\newpage

\newpage

\begin{thebibliography}{10}

\bibitem{Edmonds96}
A.~R. Edmonds, {\em Angular Momentum in Quantum Mechanics}, Princeton
  Univ.~Press, Princeton, N.J., 1996.

\bibitem{Dahlen+98}
F.~A. Dahlen and J.~Tromp, {\em Theoretical Global Seismology}, Princeton
  Univ.~Press, Princeton, N.~J., 1998.

\bibitem{Simons+2006a}
F.~J. Simons, F.~A. Dahlen, and M.~A. Wieczorek, ``Spatiospectral concentration
  on a sphere,'' {\em SIAM Rev.}~{\bf 48}(3), pp.~504--536, doi:
  10.1137/S0036144504445765, 2006.

\bibitem{Simons+2007}
F.~J. Simons and F.~A. Dahlen, ``A spatiospectral localization approach to
  estimating potential fields on the surface of a sphere from noisy, incomplete
  data taken at satellite altitudes,'' in {\em Wavelets {XII}},  D.~{Van de
  Ville}, V.~K. Goyal, and M.~Papadakis, eds.,  {\bf 6701}, pp.~670117, doi:
  10.1117/12.732406, Proc.~SPIE, 2007.

\bibitem{Simons+2006b}
F.~J. Simons and F.~A. Dahlen, ``Spherical {S}lepian functions and the polar
  gap in geodesy,'' {\em Geophys.~J.~Int.}~{\bf 166}, pp.~1039--1061, doi:
  10.1111/j.1365--246X.2006.03065.x, 2006.

\bibitem{Slepian+61}
D.~Slepian and H.~O. Pollak, ``Prolate spheroidal wave functions, {F}ourier
  analysis and uncertainty --- {I},'' {\em Bell Syst.~Tech.~J.}~{\bf 40}(1),
  pp.~43--63, 1961.

\bibitem{Landau+61}
H.~J. Landau and H.~O. Pollak, ``Prolate spheroidal wave functions, {F}ourier
  analysis and uncertainty --- {II},'' {\em Bell Syst.~Tech.~J.}~{\bf 40}(1),
  pp.~65--84, 1961.

\bibitem{Slepian64}
D.~Slepian, ``Prolate spheroidal wave functions, {F}ourier analysis and
  uncertainty --- {IV}: {E}xtensions to many dimensions; generalized prolate
  spheroidal functions,'' {\em Bell Syst.~Tech.~J.}~{\bf 43}(6),
  pp.~3009--3057, 1964.

\bibitem{Wieczorek+2005}
M.~A. Wieczorek and F.~J. Simons, ``Localized spectral analysis on the
  sphere,'' {\em Geophys.~J.~Int.}~{\bf 162}(3), pp.~655--675, doi:
  10.1111/j.1365--246X.2005.02687.x, 2005.

\bibitem{Dahlen+2008}
F.~A. Dahlen and F.~J. Simons, ``Spectral estimation on a sphere in geophysics
  and cosmology,'' {\em Geophys.~J.~Int.}~{\bf 174}, pp.~774--807, doi:
  10.1111/j.1365--246X.2008.03854.x, 2008.

\bibitem{Cox+74}
D.~R. Cox and D.~V. Hinkley, {\em Theoretical Statistics}, Chapman and Hall,
  London, UK, 1974.

\bibitem{Bendat+2000}
J.~S. Bendat and A.~G. Piersol, {\em Random data: {A}nalysis {a}nd Measurement
  Procedures}, John Wiley, New York, 3rd~ed., 2000.

\bibitem{Xu92a}
P.~Xu, ``Determination of surface gravity anomalies using gradiometric
  observables,'' {\em Geophys.~J.~Int.}~{\bf 110}, pp.~321--332, 1992.

\bibitem{Sneeuw+97}
N.~Sneeuw and M.~van Gelderen, ``The polar gap,'' in {\em Geodetic boundary
  value problems in view of the one centimeter geoid},  F.~Sans{\`o} and
  R.~Rummel, eds., {\em Lecture Notes in Earth Sciences} {\bf 65},
  pp.~559--568, Springer, Berlin, 1997.

\bibitem{Han+2008a}
S.-C. Han and F.~J. Simons, ``Spatiospectral localization of global
  geopotential fields from the {G}ravity {R}ecovery and {C}limate {E}xperiment
  {GRACE} reveals the coseismic gravity change owing to the 2004
  {S}umatra-{A}ndaman earthquake,'' {\em J.~Geophys.~Res.}~{\bf 113},
  pp.~B01405, doi: 10.1029/2007JB004927, 2008.

\bibitem{Kaula67a}
W.~M. Kaula, ``Theory of statistical analysis of data distributed over a
  sphere,'' {\em Rev.~Geophys.}~{\bf 5}(1), pp.~83--107, 1967.

\bibitem{Albertella+99}
A.~Albertella, F.~Sans{\`o}, and N.~Sneeuw, ``Band-limited functions on a
  bounded spherical domain: the {S}lepian problem on the sphere,'' {\em
  J.~Geodesy}~{\bf 73}, pp.~436--447, 1999.

\bibitem{Pail+2001}
R.~Pail, G.~Plank, and W.-D. Schuh, ``Spatially restricted data distributions
  on the sphere: the method of orthonormalized functions and applications,''
  {\em J.~Geodesy}~{\bf 75}, pp.~44--56, 2001.

\bibitem{VanGelderen+97}
M.~van Gelderen and R.~Koop, ``The use of degree variances in satellite
  gradiometry,'' {\em J.~Geodesy}~{\bf 71}, pp.~337--343, 1997.

\bibitem{Maus+2006}
S.~Maus, M.~Rother, C.~Stolle, W.~Mai, S.~Choi, H.~L{\"u}hr, D.~Cooke, and
  C.~Roth, ``Third generation of the {P}otsdam {M}agnetic {M}odel of the
  {E}arth ({POMME}),'' {\em Geochem.~Geophys.~Geosys.}~{\bf 7}, pp.~Q07008,
  doi: 10.1029/2006GC001269, 2006.

\bibitem{Gubbins+2007}
D.~Gubbins and E.~Herrero-Bervera, eds., {\em Encyclopedia of Geomagnetism and
  Paleomagnetism}, Springer, Dordrecht, Neth., 2007.

\bibitem{Chao+87}
B.~F. Chao and R.~S. Gross, ``Changes in the {E}arth's rotation and low-degree
  gravitational field induced by earthquakes,'' {\em Geophys.~J.~Int.}~{\bf
  91}, pp.~569--596, 1987.

\bibitem{Tapley+2004a}
B.~D. Tapley, S.~Bettadpur, J.~C. Ries, P.~F. Thompson, and M.~M. Watkins,
  ``{GRACE} measurements of mass variability in the {E}arth system,'' {\em
  Science}~{\bf 305}(5683), pp.~503--505, doi: 10.1126/science.1099192, 2004.

\bibitem{Han+2006b}
S.-C. Han, C.~K. Shum, M.~Bevis, C.~Ji, and C.-Y. Kuo, ``Crustal dilatation
  observed by {GRACE} after the 2004 {S}umatra-{A}ndaman earthquake,'' {\em
  Science}~{\bf 313}, pp.~658--662, doi: 10.1126/science.1128661, 2006.

\bibitem{Han+2008c}
S.-C. Han, J.~Sauber, S.~B. Luthcke, C.~Ji, and F.~F. Pollitz, ``Implications
  of postseismic gravity change following the great 2004 {S}umatra-{A}ndaman
  earthquake from the regional harmonic analysis of {GRACE} inter-satellite
  tracking data,'' {\em J.~Geophys.~Res.}~{\bf 113}, pp.~B11413, doi:
  10.1029/2008JB005705, 2008.

\bibitem{Linage+2009}
C.~de~Linage, L.~Rivera, J.~Hinderer, J.-P. Boy, Y.~Rogister, S.~Lambotte, and
  R.~Biancale, ``Separation of coseismic and postseismic gravity changes for
  the 2004 {S}umatra-{A}ndaman earthquake from 4.6 years of {GRACE}
  observations and modeling of the coseismic change by normal-modes
  summation,'' {\em Geophys.~J.~Int.}~{\bf 176}, pp.~695--714, doi:
  10.1111/j.1365--246X.2008.04025.x, 2009.

\bibitem{Dziewonski+81a}
A.~M. Dziewo{\'n}ski and D.~L. Anderson, ``Preliminary {R}eference {E}arth
  {M}odel,'' {\em Phys.~Earth Planet.~Inter.}~{\bf 25}, pp.~297--356, 1981.

\bibitem{Nissen-Meyer+2007a}
T.~Nissen-Meyer, F.~A. Dahlen, and A.~Fournier, ``Spherical-earth {F}r{\'e}chet
  sensitivity kernels,'' {\em Geophys.~J.~Int.}~{\bf 168}(3), pp.~1051--1066,
  doi: 10.1111/j.1365--246X.2006.03123.x, 2007.

\bibitem{Han+2007}
S.-C. Han and P.~Ditmar, ``Localized spectral analysis of global satellite
  gravity fields for recovering time-variable mass redistributions,'' {\em
  J.~Geodesy}~{\bf 82}(7), pp.~423--430, doi: 10.1007/s00190--007--0194--5,
  2007.

\bibitem{Han2008}
S.-C. Han, ``Improved regional gravity fields on the {M}oon from {L}unar
  {P}rospector tracking data by means of localized spherical harmonic
  functions,'' {\em J.~Geophys.~Res.}~{\bf 113}, pp.~E11012,
  doi:10.1029/2008JE003166, 2008.

\bibitem{Risbo96}
T.~Risbo, ``Fourier transform summation of {L}egendre series and
  {D}-functions,'' {\em J.~Geodesy}~{\bf 70}(7), pp.~383--396, 1996.

\bibitem{Choi+99}
C.~H. Choi, J.~Ivanic, M.~S. Gordon, and K.~Ruedenberg, ``Rapid and stable
  determination of rotation matrices between spherical harmonics by direct
  recursion,'' {\em J.~Chem.~Phys.}~{\bf 111}(19), pp.~8825--8831, 1999.

\bibitem{Wandelt+2001b}
B.~D. Wandelt and K.~M. G{\'o}rski, ``Fast convolution on the sphere,'' {\em
  Phys.~Rev.\ D}~{\bf 63}(12), p.~123002, 2001.

\bibitem{McEwen+2007}
J.~D. McEwen, M.~P. Hobson, D.~J. Mortlock, and A.~N. Lasenby, ``Fast
  directional continuous spherical wavelet transform algorithms,'' {\em IEEE
  Trans.~Signal~Process.}~{\bf 55}(2), pp.~520--529, 2007.

\bibitem{Lay+2005}
T.~Lay, H.~Kanamori, C.~J. Ammon, M.~Nettles, S.~N. Ward, R.~C. Aster, S.~L.
  Beck, S.~L. Bilek, M.~R. Brudzinski, R.~Butler, H.~R. DeShon, G.~Ekstr\"om,
  K.~Satake, and S.~Sipkin, ``The {G}reat {S}umatra-{A}ndaman earthquake of 26
  {D}ecember 2004,'' {\em Science}~{\bf 308}(5725), pp.~1127--1133, 2005.

\end{thebibliography}
\end{document}